\documentclass[showpacs,aps,floats,epsf,psfig,prc,twocolumn,graphics]{revtex4-1}
\usepackage{graphicx}
\usepackage{amssymb}
\usepackage{subfigure}
\newcommand{\isotope}[2]{$^{#2}{\rm #1}$}

\begin{document}
\title{Resolving the Reactor Neutrino Anomaly with the KATRIN Neutrino Experiment}
\author{J.~A.~Formaggio and J. Barrett}
\affiliation{Laboratory for Nuclear Science,\\
Massachusetts Institute of Technology, \\
Cambridge, MA 02139}
\email{josephf@mit.edu}
\date{\today}
\begin{abstract}
The KArlsruhe TRItium Neutrino experiment (KATRIN) combines an ultra-luminous molecular tritium source with an integrating high-resolution spectrometer to gain sensitivity to the absolute mass scale of neutrinos. The projected sensitivity of the experiment on the electron neutrino mass is 200 meV at 90\% C.L.  With such unprecedented resolution, the experiment is also sensitive to physics beyond the Standard Model, particularly to the existence of additional sterile neutrinos at the eV mass scale. A recent analysis of available reactor data appears to favor the existence of such such a sterile neutrino with a mass splitting of $|\Delta m_{\rm sterile}|^2 \ge 1.5$ eV$^2$ and mixing strength of $\sin^2{2\theta_{\rm sterile}} = 0.17\pm 0.08$ at 95\% C.L.  Upcoming tritium beta decay experiments should be able to rule out or confirm the presence of the new phenomenon for a substantial fraction of the allowed parameter space.
\end{abstract}
\maketitle

\paragraph{Introduction \label{sec:Introduction}} 

The theory of neutrino oscillations, as described by the Pontecorvo-Maki-Nakagawa-Sakata (PMNS) mixing matrix, provides a simple and well-grounded description of the neutrino data obtained thus far~\cite{Nakamura:2010zzi}.  Neutrino oscillation experiments carried out over the past half-century firmly establish the presence of the phenomena to the point that the existence of non-zero neutrino masses is no longer in question.  Current experiments are now engaged in gaining greater precision of the relevant mixing and mass parameters in a manner similar to that which was done for the quark sector.

As the precision of such experiments continues to improve and the tools by which the data is analyzed increase in sophistication, one finds that the emerging picture may be more complex than previously realized.  A recent re-analysis of existing reactor data by Mention and collaborators~\cite{Mention:2011rk} appears consistent with the presence of a fourth, sterile neutrino.  A combined analysis using available reactor data, as well as data collected by gallium solar neutrino calibration experiments~\cite{Anselmann:1995ag,Abdurashitov:2005tb} and the MiniBooNE neutrino data~\cite{AguilarArevalo:2007it} leads to a mass splitting of $|\Delta m_{\rm sterile}^2| > 1.5$ eV$^2$ and $\sin{(2\theta_{\rm s})}^2 = 0.17 \pm 0.08$ at 95\% C.L.  The observation, herein referred to as ``the reactor neutrino anomaly," appears to be further collaborated by other observations, particularly the more recent data collected by the MiniBooNE experiment in antineutrino running~\cite{AguilarArevalo:2010wv}.   At this stage it is too early to make any strong claim as to the validity of one or all of these observations.  Continued data collection and scrutiny of systematic uncertainties will provide better guidance as to whether new physics is at play.

The existence of a sterile neutrino with an eV mass scale would certainly have a profound impact on our understanding of physics beyond the Standard Model.  However, it should be noted that such new physics would also have a profound {\em practical} impact on the current experimental research program, particularly for upcoming long-baseline experiments.  Experiments dedicated to measuring the level of CP violation in the neutrino sector, such as the Fermilab Long-Baseline Neutrino Experiment (LBNE)~\cite{bib:LBNE}, could be seriously be impacted by the presence of a eV-scale sterile neutrino with moderate mixing to the Standard Model neutrinos.  Such mixing would occur at a baseline similar to that of the near detector, potentially affecting the flux normalization of the experiment.  Should such non-Standard Model neutrinos behave differently than their anti-neutrino counterparts, it may further confound any collected data.  Resolution of the reactor neutrino anomaly should take place quickly and be pursued with whatever existing tools may be at the community's disposal.

In this paper, we propose that the KArlsruhe TRItium Neutrino mass experiment (KATRIN) is well poised to resolve the question whether the reactor neutrino anomaly can be interpreted as the existence of a light sterile neutrino.  Given the projected sensitivity of the KATRIN experiment, the anomaly should provide a definite signal in its energy spectrum.  The experiment also provides a check to the anomaly that is orthogonal to data obtained from reactor and short baseline experiments.  In this paper, we discuss the potential sensitivity of the KATRIN experiment within the context of the observed reactor anomaly.


\paragraph{Sterile Neutrino Signatures in Beta Decay}
\label{sec:signal}

The most sensitive direct searches for the electron neutrino mass
up to now are based on the investigation of the electron spectrum
of tritium $\beta$-decay.  In the presence of mixing, the electron neutrino is a combination of the mass eigenstates $\nu_i$ with masses $m_i$ such that $\nu_e = \sum_i U_{ei} \nu_i$.  The corresponding electron energy spectrum is given by

\begin{widetext}
\begin{eqnarray*}
\label{eq:decay}
{dN \over dK_e} = N_{\rm decay} \cdot F(Z,K_e) \cdot p_e \cdot (K_e+m_e) \cdot (E_0-K_e) \cdot \sum_{i=1,3}|U_{ei}|^2\sqrt{(E_0-K_e)^2-m_i^2} \cdot \Theta(E_0 - K_e - m_i).
\end{eqnarray*}
\end{widetext}
 
\noindent where $K_e$ denotes the electron kinetic energy, $p_e$ is the electron
momentum, $m_e$ is the electron mass, $E_0$ is the endpoint energy of the $^3$H$_2 \rightarrow (^3$He$^3$H)$^+ + e^- +\bar{\nu}_e$ decay, $N_{\rm decay}$ is the rate of tritium beta decay, $F(Z,K_e)$ is the Fermi function, taking into account the Coulomb interaction of the outgoing electron in the final state, $Z$ is the atomic number of the final state, $\Theta$ is the step function imposed by energy conservation, and $U_{ei}$ is the element from the PMNS mixing matrix. As both the matrix elements and $F(Z,K_e)$ are independent of the neutrino mass, the dependence of the spectral shape on $m_i$ is given solely by the phase space factor. In addition, the bound on the neutrino mass from tritium $\beta$-decay is independent of whether the electron neutrino is a Majorana or a Dirac particle.

It is possible to define an effective kinematic mass term, $m_\beta$, which is the incoherent sum of the neutrino masses, $m^2_\beta = \sum_{i=1}^3 |U_{ei}|^2 m_i^2$.  As a result, the electron spectrum becomes dependent upon a single, effective mass parameter. This approximation works well when the mass splittings are well below the resolution achievable by the experiment~\cite{Farzan2003224}. 

The reactor neutrino anomaly seems to indicate the possibility of a fourth sterile neutrino with a small (non-zero) mixing and relatively large mass splitting, at least compared to the active neutrinos.  We can reformulate the above expression to better highlight the presence of such a splitting.  Let us define an average lower and upper mass regime, $\bar{m}_L$ and $\bar{m}_U$, as follows:

\begin{eqnarray}
\bar{m}_L^2 = \frac{\sum_{i=1}^{N_L} |U_{ei}|^2 m_i^2}{\sum_{i=1}^{N_L} |U_{ei}|^2};\\
\bar{m}_U^2 = \frac{\sum_{i=N_L+1}^{N} |U_{ei}|^2 m_i^2}{\sum_{i=N_L+1}^{N} |U_{ei}|^2},
\end{eqnarray}

\noindent where $N_L$ is the number of neutrino masses on the lower mass scale and $N$ is the total number of neutrino species.  Taking advantage of the unitarity condition, $\sum_{i=1}^{N} |U_{ei}|^2 = 1$, and letting $|U_S|^2 = \sum_{i=N_L+1}^{N} |U_{ei}|^2$, we can re-write the decay phase as:

\begin{widetext}
\begin{eqnarray}
\label{eq:Us}
\phi(|U_S|^2, \bar{m}_L, \bar{m}_U) \simeq z^2 \left( (1-|U_S|^2)\sqrt{1-\frac{\bar{m}_L^2}{z^2}} \cdot \Theta(z-\bar{m}_L) + |U_S|^2\sqrt{1-\frac{\bar{m}_U^2}{z^2}} \cdot \Theta(z-\bar{m}_U)\right)
\end{eqnarray}
\end{widetext}

\noindent where $z = (E_0 - K_e)$. We consider just the case where only one large splitting is present whose scale is dictated by $m_S^2 = \bar{m}_U^2 - \bar{m}_L^2 \simeq \Delta m^2_S$.  Note that this formulation is relatively insensitive to the details of the splitting and ordering of the mass spectrum, as long as the smaller splittings are below the resolution of the experiment.   For the case of one single sterile neutrino, such as posited by 3+1 models~\cite{bib:deGouvea}, it is possible to express the amplitude $|U_S|^2$ in terms of a mixing angle analogous to that of ordinary neutrino mixings:

\begin{equation}
\sin^2{2\theta_s} = 4 |U_S|^2 (1-|U_S|^2)
\end{equation}

The mixing angle $\sin^2{2\theta_s}$ mirrors that employed by Mention {\it et al.}, though it should be stressed that tritium beta decay experiments are primarily sensitive to the effective mass splitting.

\paragraph{The KATRIN Beta Decay Experiment}
\label{sec:KATRIN}

In this paper, we primarily focus on the sensitivity of the KATRIN neutrino mass experiment to the reactor neutrino anomaly.  The KATRIN experiment is the next generation tritium beta decay experiment with sub-eV sensitivity to make a direct, model-independent measurement of the electron neutrino mass. The principle of the experiment is to look for a distortion at the high energy endpoint of the electron spectrum of tritium $\beta$-decay.  The shape of the electron energy spectrum of tritium beta decay is determined by well-understood and measurable quantities. Any deviation from this shape would be directly attributable to neutrino mass and would allow a direct determination of the mass of the electron neutrino. After three years of running and in the absence of a positive signal, KATRIN can place a limit on the neutrino mass below 200 meV at the 90\% C.L.  This level represents an order of magnitude improvement in the measured limit on the absolute neutrino mass scale. 

The experiment is located at the Karlsruhe Institute of Technology (KIT). This is a unique location as it hosts the Tritium Laboratory Karlsruhe, which is in charge of the tritium cycle of the international ITER fusion program. KATRIN will use the windowless gaseous tritium source technique, as used by Los Alamos~\cite{robertson1991lea} and Troitsk~\cite{lobashev2001dsn}. Decay electrons from the source pass through a 10-meter long differential and cryogenic pumping subsection guided by superconducting magnets. The purpose of the differential pumping system is to prevent gas from entering the spectrometer system, which would degrade resolution and raise background by contaminating the system with tritium. 

The KATRIN spectrometer is based on technology developed by the Mainz~\cite{kraus2005frp} and Troitsk~\cite{lobashev2001dsn} tritium beta decay experiments. These experiments used a so-called MAC-E-Filter (Magnetic Adiabatic Collimation combined with an Electrostatic filter). This technology draws the isotropic electrons from a decay or capture event along magnetic field lines through a decreasing magnetic field so that the cyclotron motion of the electrons around the magnetic field lines is transformed into longitudinal motion along the magnetic field lines. A retarding potential is applied such that only electrons with energy greater than the retarding potential are transmitted to an electron counting detector. By varying the retarding potential, the shape of the decay spectrum can be reconstructed. The energy resolution of this measurement is determined by the ratio $\frac{\Delta K_e}{K_e} = \frac{B_A}{B_{\rm max}} = \frac{1}{20000}$, where $B_A$ is the magnetic field in the analyzing plane (the point of maximum potential) and $B_{\rm max}$ is the maximum magnetic field. The decay electrons exiting the spectrometer are imaged onto a silicon PIN diode array using a 6 T superconducting magnet. The electrons counted in the detector array comprise the signal needed to reconstruct the beta-decay spectrum.  The experiment is currently in the final phases of construction.

The KATRIN experiment gains greater sensitivity to the neutrino mass scale mainly from its intense tritium source.  The number of observed tritium decays is given by \[ N_{\rm decay} = N(T_2) \cdot \frac{\Delta \Omega}{2\pi} \cdot P \] where $N(T_2)$ is the number of tritium molecules, $\frac{\Delta \Omega}{2\pi}$ is the solid angle of the source as seen by the detector, and $P$ is the probability that the emitted electron exits of the source without undergoing inelastic scattering.

The number of $T_2$ molecules in the source is given by \[ N(T_2) = \rho d \cdot \epsilon_T \cdot A_s, \] where $\rho d$ is the source column density, $\epsilon_T=0.95$ is the tritium purity, and $A_s=52.65$ cm$^2$ is the source area. Thus the signal rate can be written as \[ N_{\rm decay} = A_s \cdot \epsilon_T \cdot \frac{\Delta \Omega}{2\pi} \cdot P \cdot \rho d. \] The factor $P \cdot \rho d$ can be considered as an effective column density, $\rho d_{\rm eff}$, which has the value $2.1\times10^{17}/$cm$^2$ at KATRIN~\cite{angrik2005kdr}.  The effective tritium source strength is equivalent to $N_{\rm decay} = 3.8\times 10^{18}$ tritium molecules, or an equivalent mass of approximately $\sim 40~\mu$g.

Because the KATRIN experiment measures the integral beta decay spectrum above the retarding potential $qU$, the electron spectrum is really the convolution of the $\beta$~electron spectrum, $dN/dK_e$, and the transmission function of the detector, $T(K_e,qU)$.  KATRIN also expects a small but finite background rate, $N_b$, to contribute to the overall signal.  Currently, this background rate is expected to be of order 10 mHz in the signal region of interest, independent of retarding potential.  The observed $\beta$-decay energy spectrum is therefore:  

\begin{equation}
\label{eq:spectrum}
R(qU) = \int_{qU}^{\infty} \frac{dN}{dK_e} T(K_e,qU) dK_e + N_b.
\end{equation}

The transmission function, $T(K_e,qU)$, depends on the value of the retarding potential, $qU$,  as well as the intrinsic resolution of the main spectrometer. For an isotropic source, $T(K_e,qU)$ can be written analytically as:

\begin{equation}
	T(K_e,qU)= \left\{ \begin{array}{ll}
         0 & \mbox{if $K_e-qU <  0$}\\
	\frac{1-\sqrt{1-\frac{(K_e-qU) B_S}{K_e B_A}}}{1-\sqrt{1-\frac{\Delta K_e B_S}{K_e B_{\rm max}}}} & \mbox{if $0 \leq K_e-qU \leq \Delta K_e$}\\
         1 & \mbox{if $K_e-qU > \Delta K_e$}.\end{array} \right.
	\label{eq:transmission}
\end{equation}

\noindent where $K_e$ is the electron energy, $B_S$ is the magnetic field at the source, $B_A$ is the magnetic field at qU, $B_{\rm max}$ is the maximum (pinch) field, and $\Delta K_e = 0.93$ eV at the endpoint energy.  

Since the source involves the presence of molecular T$_2$ gas, one must include any corrections to the endpoint energy due to interactions with the molecular daughter molecule following the tritium decay. An accounting of these states is given in~\cite{saenz2000imf}.  Of most relevance are the effects of the rotational-vibrational contributions from decays to the ground state, which introduce a mean excitation energy of 1.7 eV with an inherent broadening of 0.36 eV.  In this analysis, the final states are taken into account via a summation over the states of the He$^+$T molecule, each final state weighted by the probability for that state occurring. The final states also impose a practical upper limit on the search for a large mass splitting, since for energies above 13 eV ionization losses become a significant energy loss in the system.

\paragraph{Results and Discussion}

\begin{figure}[ht]
\includegraphics[width=1.1\columnwidth]{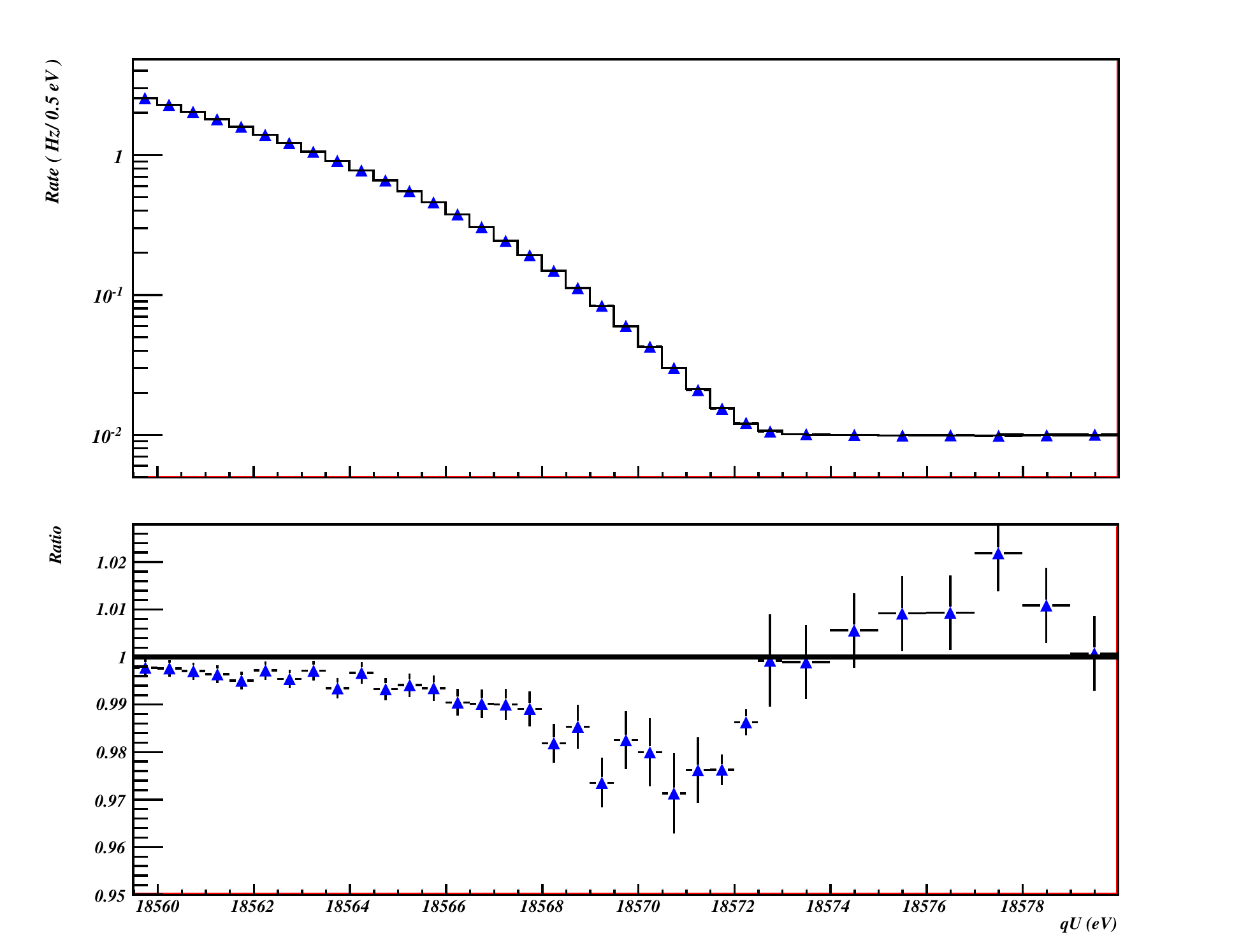} 
\caption{Convoluted three-year tritium beta decay spectrum as observed by the KATRIN experiment (top). The zero-mass, active-only spectrum (solid curve) is contrasted against a spectrum with a high mass sterile neutrino ($\bar{m}_L = 0, \bar{m}_U = 2.0$ eV, $|U_S|^2 = 0.067$; triangle markers).  Ratio between 2 eV and 0 eV spectra is shown (bottom).}
\label{fig:spectrum}
\end{figure} 

\begin{figure}[ht]
\includegraphics[width=1.25\columnwidth]{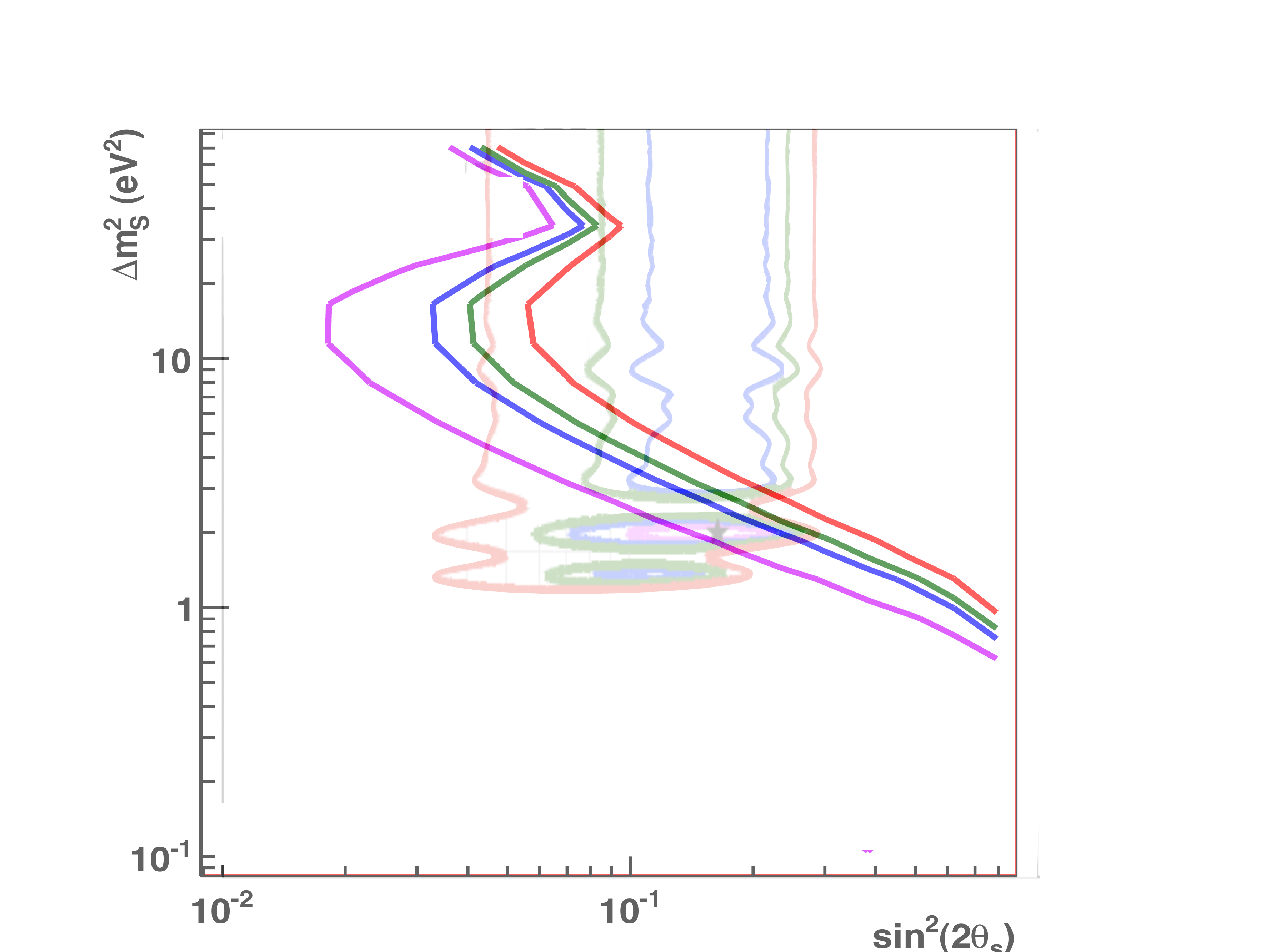} 
\caption{Sensitivity of the KATRIN neutrino mass measurement for a sterile neutrino with relatively large mass splitting (solid contours).  Figures shows exclusion curves of mixing angle $\sin^2{(2\theta_S)}$ versus mass splitting $|\Delta m_S^2|$ for the 68\% (violet), 90\% (blue), 95\% (green), and 99\% (red) C.L. after three years of data taking.  Figure 7 from Ref.~\cite{Mention:2011rk} show in the background.}
\label{fig:sensitivity}
\end{figure} 

The sensitivity of KATRIN to sterile neutrinos can be scaled directly from the experiment's sensitivity to the degenerate mass scale, $\sigma_m$; as the errors on $m_S^2$ and $|U_S|^2$ scale roughly as $\sigma_m/|U_S|^2$ and $\sigma_m/m_S^2$, respectively.  To calculate the sensitivity, however, we make use of a full simulation of the spectrum as seen by the experiment.  Figure~\ref{fig:spectrum} illustrates a sample convoluted spectrum.  The spectrum is fit to a function of the same form as Eq.~\ref{eq:Us}. An extended log-likelihood is calculated to compare a simulated 3-year KATRIN spectrum -- with systematic variations-- against a model beta decay spectrum. The data are fit from -15 eV to +5 eV from the endpoint kinetic energy, with the running time at each voltage dictated by the KATRIN optimal running scan, where the majority of the data is taken near the endpoint to provide enhanced statistical gain.  The two mass scales ($\bar{m}_U$ and $\bar{m}_L$), the sterile admixture ($|U_S|^2$), the endpoint, the background rate and the overall decay amplitude are treated as free parameters.  Dominant systematics errors such as high voltage stability, magnetic field precision, the effect of final states, and the error on the number of available tritium atoms are included in the analysis.

To determine the potential sensitivity of KATRIN to low mass sterile neutrinos, we compare the relative likelihood (${\cal L}$) between the sterile and non-sterile models:

\begin{equation}
\Delta {\cal L} = {\cal L}(U_S^2, \bar{m}_U, \bar{m}_L) - {\cal L}(m_\beta).
\end{equation}

The distortion caused by a non-zero $|U_S|^2$ is statistically distinguishable from both the zero mass case and the non-sterile scenario ($|U_S|^2 = 0$). Although the best fit island suggested by the reactor anomaly data ($\Delta m_S^2 \ge 1.5$ eV$^2$) is below the 90\% C.L. reach of KATRIN, the tritium measurement removes a substantial fraction of the allowed phase space.  KATRIN is able to place a lower limit on the mixing angle $\sin^2{(2\theta_s)} \ge 0.08$ at the 95\% C.L. for mass splitting $\Delta m_s^2 \ge 4$ eV$^2$ after three years of data taking (see Figure~\ref{fig:sensitivity}).  The final sensitivity curves are somewhat sensitive to the data running model used, with the greatest sensitivity gained in the planned optimal running scenario (versus a more uniform run time distribution).  These results are consistent with an earlier sterile neutrino sensitivity study proposed for present and next generation beta decay experiments~\cite{bib:deGouvea,Riis:2010zm}.

In summary, an observation of a kink in the beta decay spectrum, combined with existing oscillation data, would indicate strong evidence for the existence of a light sterile neutrino.  KATRIN is an relatively unique position to address the reactor neutrino anomaly. The nature of the measurement itself is essentially orthogonal to that provided by rate measurements obtained from reactor data.  The KATRIN experiment is also able to provide results on the reactor anomaly in a relatively short time scale.   Should a signal manifest itself in the data, the observation could be further confirmed by future beta decay experiments, such as MARE ~\cite{bib:MARE} and Project 8~\cite{bib:project8}.  It should be noted that if the MARE experiment were to employ the electron capture of \isotope{Ho}{163} in place of beta decay of \isotope{Re}{187}, their measurement would also be able to probe differences between neutrino and anti-neutrinos in a model-independent manner.  Other techniques, such as the study of coherent scattering using low energy intense mono-energetic neutrino sources also appears promising~\cite{bib:Formaggio}.  A more in-depth study of the reactor beta decay spectrum --essentially a follow-up to the ILL measurement program-- would also strengthen the case for the observation.

\section{Acknowledgments}

The authors would like to thank Andre de Gouvea, Georgia Karagiorgi and the MIT Neutrino and Dark Matter Group for insightful discussions on the implications of the reactor anomaly.  This work is supported by the U.S. Department of Energy under Grant No. DE-FG02-06ER-41420.


\begin{thebibliography}{18}
\expandafter\ifx\csname natexlab\endcsname\relax\def\natexlab#1{#1}\fi
\expandafter\ifx\csname bibnamefont\endcsname\relax
  \def\bibnamefont#1{#1}\fi
\expandafter\ifx\csname bibfnamefont\endcsname\relax
  \def\bibfnamefont#1{#1}\fi
\expandafter\ifx\csname citenamefont\endcsname\relax
  \def\citenamefont#1{#1}\fi
\expandafter\ifx\csname url\endcsname\relax
  \def\url#1{\texttt{#1}}\fi
\expandafter\ifx\csname urlprefix\endcsname\relax\def\urlprefix{URL }\fi
\providecommand{\bibinfo}[2]{#2}
\providecommand{\eprint}[2][]{\url{#2}}

\bibitem[{\citenamefont{Nakamura et~al.}(2010)}]{Nakamura:2010zzi}
\bibinfo{author}{\bibfnamefont{K.}~\bibnamefont{Nakamura}} \bibnamefont{et~al.}
  (\bibinfo{collaboration}{Particle Data Group}), \bibinfo{journal}{J.Phys.G}
  \textbf{\bibinfo{volume}{G37}}, \bibinfo{pages}{075021}
  (\bibinfo{year}{2010}).

\bibitem[{\citenamefont{Mention et~al.}(2011)}]{Mention:2011rk}
\bibinfo{author}{\bibfnamefont{G.}~\bibnamefont{Mention}} \bibnamefont{et~al.}
  (\bibinfo{year}{2011}), \eprint{hep-ex:1101.2755}.

\bibitem[{\citenamefont{Anselmann et~al.}(1995)}]{Anselmann:1995ag}
\bibinfo{author}{\bibfnamefont{P.}~\bibnamefont{Anselmann}}
  \bibnamefont{et~al.} (\bibinfo{collaboration}{GALLEX Collaboration}),
  \bibinfo{journal}{Phys.Lett.} \textbf{\bibinfo{volume}{B357}},
  \bibinfo{pages}{237} (\bibinfo{year}{1995}).

\bibitem[{\citenamefont{Abdurashitov et~al.}(2006)}]{Abdurashitov:2005tb}
\bibinfo{author}{\bibfnamefont{J.}~\bibnamefont{Abdurashitov}}
  \bibnamefont{et~al.}, \bibinfo{journal}{Phys.Rev.}
  \textbf{\bibinfo{volume}{C73}}, \bibinfo{pages}{045805}
  (\bibinfo{year}{2006}).

\bibitem[{\citenamefont{Aguilar-Arevalo et~al.}(2007)}]{AguilarArevalo:2007it}
\bibinfo{author}{\bibfnamefont{A.}~\bibnamefont{Aguilar-Arevalo}}
  \bibnamefont{et~al.} (\bibinfo{collaboration}{The MiniBooNE Collaboration}),
  \bibinfo{journal}{Phys.Rev.Lett.} \textbf{\bibinfo{volume}{98}},
  \bibinfo{pages}{231801} (\bibinfo{year}{2007}), \eprint{0704.1500}.

\bibitem[{\citenamefont{Aguilar-Arevalo et~al.}(2010)}]{AguilarArevalo:2010wv}
\bibinfo{author}{\bibfnamefont{A.}~\bibnamefont{Aguilar-Arevalo}}
  \bibnamefont{et~al.} (\bibinfo{collaboration}{The MiniBooNE Collaboration}),
  \bibinfo{journal}{Phys.Rev.Lett.} \textbf{\bibinfo{volume}{105}},
  \bibinfo{pages}{181801} (\bibinfo{year}{2010}), \eprint{1007.1150}.

\bibitem[{\citenamefont{Maricic}(2010)}]{bib:LBNE}
\bibinfo{author}{\bibfnamefont{J.}~\bibnamefont{Maricic}}
  (\bibinfo{collaboration}{LBNE DUSEL Collaboration}),
  \bibinfo{journal}{J.Phys.Conf.Ser.} \textbf{\bibinfo{volume}{203}},
  \bibinfo{pages}{012109} (\bibinfo{year}{2010}).

\bibitem[{\citenamefont{Farzan and Smirnov}(2003)}]{Farzan2003224}
\bibinfo{author}{\bibfnamefont{Y.}~\bibnamefont{Farzan}} \bibnamefont{and}
  \bibinfo{author}{\bibfnamefont{A.~Y.} \bibnamefont{Smirnov}},
  \bibinfo{journal}{Physics Letters B} \textbf{\bibinfo{volume}{557}},
  \bibinfo{pages}{224 } (\bibinfo{year}{2003}), ISSN \bibinfo{issn}{0370-2693}.

\bibitem[{\citenamefont{de~Gouvea et~al.}(2007)}]{bib:deGouvea}
\bibinfo{author}{\bibfnamefont{A.}~\bibnamefont{de~Gouvea}}
  \bibnamefont{et~al.}, \bibinfo{journal}{Phys.Rev.}
  \textbf{\bibinfo{volume}{D75}}, \bibinfo{pages}{013003}
  (\bibinfo{year}{2007}).

\bibitem[{\citenamefont{Robertson et~al.}(1991)}]{robertson1991lea}
\bibinfo{author}{\bibfnamefont{R.}~\bibnamefont{Robertson}}
  \bibnamefont{et~al.}, \bibinfo{journal}{Physical Review Letters}
  \textbf{\bibinfo{volume}{67}} (\bibinfo{year}{1991}).

\bibitem[{\citenamefont{Lobashev et~al.}(2001)}]{lobashev2001dsn}
\bibinfo{author}{\bibfnamefont{V.}~\bibnamefont{Lobashev}}
  \bibnamefont{et~al.}, \bibinfo{journal}{Nuclear Physics-Section
  B-PS-Proceedings Supplements} \textbf{\bibinfo{volume}{91}},
  \bibinfo{pages}{280} (\bibinfo{year}{2001}).

\bibitem[{\citenamefont{Kraus et~al.}(2005)}]{kraus2005frp}
\bibinfo{author}{\bibfnamefont{C.}~\bibnamefont{Kraus}} \bibnamefont{et~al.},
  \bibinfo{journal}{The European Physical Journal C-Particles and Fields}
  \textbf{\bibinfo{volume}{40}}, \bibinfo{pages}{447} (\bibinfo{year}{2005}).

\bibitem[{\citenamefont{Angrik et~al.}(2005)}]{angrik2005kdr}
\bibinfo{author}{\bibfnamefont{J.}~\bibnamefont{Angrik}} \bibnamefont{et~al.},
  \bibinfo{journal}{Wissenschaftliche Berichte FZKA}
  \textbf{\bibinfo{volume}{7090}} (\bibinfo{year}{2005}).

\bibitem[{\citenamefont{Saenz et~al.}(2000)\citenamefont{Saenz, Jonsell, and
  Froelich}}]{saenz2000imf}
\bibinfo{author}{\bibfnamefont{A.}~\bibnamefont{Saenz}},
  \bibinfo{author}{\bibfnamefont{S.}~\bibnamefont{Jonsell}}, \bibnamefont{and}
  \bibinfo{author}{\bibfnamefont{P.}~\bibnamefont{Froelich}},
  \bibinfo{journal}{Phys Rev Lett} \textbf{\bibinfo{volume}{84}},
  \bibinfo{pages}{242} (\bibinfo{year}{2000}).

\bibitem[{\citenamefont{Riis and Hannestad}(2011)}]{Riis:2010zm}
\bibinfo{author}{\bibfnamefont{A.~S.} \bibnamefont{Riis}} \bibnamefont{and}
  \bibinfo{author}{\bibfnamefont{S.}~\bibnamefont{Hannestad}},
  \bibinfo{journal}{JCAP} \textbf{\bibinfo{volume}{1102}}, \bibinfo{pages}{011}
  (\bibinfo{year}{2011}), \eprint{1008.1495}.

\bibitem[{\citenamefont{Alessandrello et~al.}(1999)}]{bib:MARE}
\bibinfo{author}{\bibfnamefont{A.}~\bibnamefont{Alessandrello}}
  \bibnamefont{et~al.}, \bibinfo{journal}{Phys. Rev. Lett.}
  \textbf{\bibinfo{volume}{82}}, \bibinfo{pages}{513} (\bibinfo{year}{1999}).

\bibitem[{\citenamefont{Monreal and Formaggio}(2009)}]{bib:project8}
\bibinfo{author}{\bibfnamefont{B.}~\bibnamefont{Monreal}} \bibnamefont{and}
  \bibinfo{author}{\bibfnamefont{J.}~\bibnamefont{Formaggio}},
  \bibinfo{journal}{Phys. Rev. D} \textbf{\bibinfo{volume}{80}},
  \bibinfo{pages}{051301} (\bibinfo{year}{2009}).

\bibitem[{\citenamefont{Formaggio}(2011)}]{bib:Formaggio}
\bibinfo{author}{\bibfnamefont{J.~A.} \bibnamefont{Formaggio}}
  (\bibinfo{year}{2011}), \bibinfo{note}{article in preparation}.

\end{thebibliography}
\end{document}